\documentclass[runningheads]{llncs}
\usepackage[T1]{fontenc}
\usepackage{graphicx}
\usepackage{amsmath}
\usepackage{amsfonts}
\usepackage{amssymb}
\usepackage{xcolor}
\usepackage{comment}

\usepackage{array}
\usepackage{bbding}

\newcommand{\la}{\langle}
\newcommand{\ra}{\rangle}

\begin{document}
\def\mi#1{\mathit{#1}}
\title{The Interplay Between High-Level Problems And The Process Instances That Give Rise To Them\thanks{We thank the Alexander von Humboldt (AvH) Stiftung for supporting our research.}}
%The Interplay between process dynamics and the cases that give rise to it
%
\titlerunning{High-level Problems and Their Underlying Process Instances}
\author{Bianka Bakullari\,\Envelope\,\inst{1}\orcidID{0000-0003-2680-0826} \and
Jules van Thoor\inst{2} \and \\
Dirk Fahland\inst{2}\orcidID{0000-0002-1993-9363} \and
Wil van der Aalst \inst{1}\orcidID{0000-0002-0955-6940}}
\authorrunning{B. Bakullari et al.}
\institute{RWTH Aachen University, Germany \and
Eindhoven University of Technology, Netherlands\\
\email{\{bianka.bakullari, wvdaalst\}@pads.rwth-aachen.de}\\
\email{d.fahland@tue.nl }}
\maketitle              % typeset the header of the contribution
\begin{abstract}
Business processes may face a variety of problems due to the number of tasks that need to be handled within short time periods, resources' workload and working patterns, as well as bottlenecks.
These problems may arise locally and be short-lived, but as the process is forced to operate outside its standard capacity, the effect on the underlying process instances can be costly.
We use the term \emph{high-level behavior} to cover all process behavior which can not be captured in terms of the individual process instances.
%Whenever such behavior emerges, we call the cases which are involved in it \emph{participating cases}.
The natural question arises as to how the characteristics of cases relate to the high-level behavior they give rise to.
In this work, we first show how to detect and correlate observations of high-level problems, as well as determine the corresponding (non-)participating cases.
Then we show how to assess the connection between any case-level characteristic and any given detected sequence of high-level problems.
Applying our method on the event data of a real loan application process revealed which specific combinations of delays, batching and busy resources at which particular parts of the process correlate with an application's duration and chance of a positive outcome. 

\end{abstract}

\section{Introduction}\label{sec:introduction}
\begin{figure}[t]
\centerline{\includegraphics[scale=0.52]{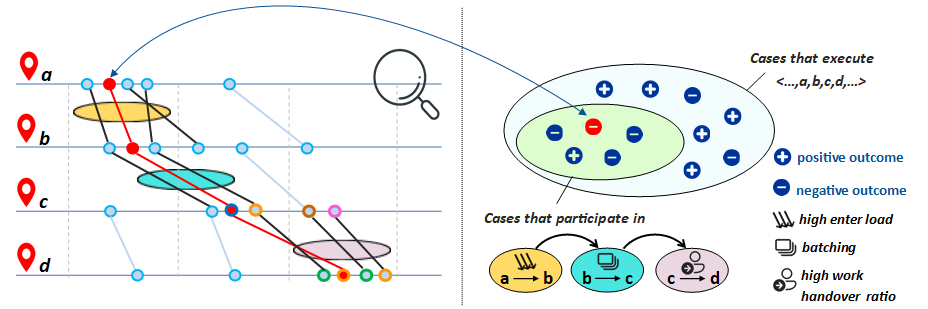}}
\caption{An illustration of the approach: 
On the left, a sketch of the performance spectrum \cite{perf-spectrum} showing process instances running through segments $(a,b)$, $(b,c)$ and $(c,d)$.
At each segment, several event pairs give rise to various patterns: high entering load at $(a,b)$ (top cloud), batching at $(b,c)$ (middle cloud), and high work handover ratio at $(c,d)$ (bottom cloud).
The cases which are involved in this pattern sequence are considered as ``participating" w.r.t. that sequence.
Given a case-level property, we analyze how its value changes when comparing these cases (e.g., the process outcome of the participating red case is negative) to the cases which visited the same locations where the pattern emerged (here $(a,b)$, $(b,c)$, and $(c,d)$), but did not give rise to such pattern.}
\label{fig: intro}
\end{figure}
\subsection{Motivation}\label{sec:motivation}
Process mining techniques analyze event data stored in information systems in order to get insights of real business processes \cite{pm}.
Organizations strive to improve their running processes by reducing cost and waste, improving resource utilization and customer satisfaction, and so on.
Many Key Performance Indicators (KPIs) describe the process in terms of the individual process instances (also called \emph{cases}), e.g., by referring to the average time it takes for a case to complete the process (the \emph{throughput time}), the average accumulated cost or positive outcome rate.
Process instances, however, do not run in isolation.
From this viewpoint, cases that are simultaneously active in a process resemble cars moving along traffic.
Cars can cause traffic jams which, in turn, cause delays and accidents.
Similarly, cases may overload the process and the workers, leading to congestion and delays.
Moreover, when attending to multiple active cases, resources may execute work in \emph{batches} which, in turn, also influences the manner in which process traffic moves forward.
We refer to this kind of emergent process behavior, which is not detectable at the level of the individual instances, as \emph{high-level behavior}.
This behavior is \emph{dynamic}; that is, it may arise locally and be short-lived, but it can have an influence on the process runs of the cases active at that time.
Nevertheless, similar to traffic jams, there is always a specific set of cases involved whenever such behavior emerges.
On one hand, the characteristics of a case may aggravate the emergence of high-level behavior, e.g., a demanding case can block resources for longer time periods. 
On the other hand, the outcome of a case can also be affected by high-level behavior occurring throughout its run, e.g., a case may not receive the necessary attention if by chance it enters the process in a busy period.
There is obviously an interplay between the high-level behavior that arises in the process and the cases which give rise to it. 
Our method explores this interplay by detecting which patterns of high-level behavior emerge surprisingly often from specific case types.
The advantage of possessing this knowledge can be manifold.
Depending on the case property at hand, one can adjust the process for specific types of cases in order to avoid undesired but expected high-level problems, or one can make a better online prediction of the progress of a case given its involvement in specific patterns of high-level behavior.
\subsection{Example}\label{sec:example}
Suppose that the red lines in Fig. \ref{fig: intro} describe the process run of a case which executes activities $\la a,b,c,d \ra$.
Assume that, in this process, it is unusual to observe four cases entering segment $(a,b)$ simultaneously in a short time frame (the lines within the top yellow cloud). 
Moreover, multiple cases including the red case execute activities $b$ and $c$ very close to each other with a very similar waiting time in-between (the lines within the middle blue cloud).
Later in the process, four cases traverse segment $(c,d)$ in a short time frame (the lines within the bottom violet cloud) and work is handed over from four resources (four different colors encircling the $c$ events) to only two resources (only green and orange encircling the four $d$ events).
Assuming that this work handover ratio is unusually high, we can claim that in this example, the red case is involved in three patterns at three different process segments it traverses: \emph{high enter load} at $(a,b)$, \emph{batching} at $(b,c)$ and \emph{high work handover ratio} at $(c,d)$.
Now suppose that the red case turns out to have a negative outcome in the process (see the red case's minus sign and positioning in the right side of Fig. \ref{fig: intro}).
The question arises whether participating in this specific sequence of high-level patterns increases or decreases the likelihood of a negative outcome in the process.
In this study, we evaluate whether a particular type of cases is disproportionately represented within the case set that generates a specific episode of high-level behavior.
When such a situation occurs, we consider the connection between that specific episode and case property as a valuable insight into the behavior of the process.
\subsection{Approach}\label{sec:approach}
As shown in the previous example, diverse types of high-level patterns can emerge at different locations and times within the process.
To be able to compute how strongly cases' participation in high-level behavior correlates with any particular case characteristic, we first need to define what high-level behavior may look like (the clouds in Fig. \ref{fig: intro}).
In this work, we introduce different types of high-level behavior at the segment level that relate to load (enter and exit rates), resource busyness (handover ratio and workload), and working patterns (batches and delays).
We use the idea introduced in previous work \cite{hlem} and conceptualize each outlier observation of such behavior using \emph{high-level events}.
It is worth emphasizing that multiple concurrently active cases can give rise to various forms of high-level behavior.
This behavior can refer not only to process traffic and workload but also e.g., compliance with regulations which guide how the process should be executed given any particular set of active cases.
Within this work, we outline high-level behavior that is specifically related to congestion as it is commonly observed across different process domains.
The method could, however, be easily extended to any other type of high-level problem that arises from a set of cases that pass through a process segment in close time proximity.
Given a high-level event, we determine what qualifies a case as ``(non-)participating'' (where to assign each case w.r.t. the sets depicted as circles in the right side of Fig. \ref{fig: intro}). 
Any two high-level events with sufficient overlap in time, location and underlying case sets are assumed to be correlated, leading this way to sequences of subsequently connected high-level events (such as the sequence consisting of the three clouds in Fig. \ref{fig: intro}).
A case participates in such a \emph{high-level path} whenever its events are involved in each high-level event comprising the path.
In Fig. \ref{fig: intro}, these are the cases whose black lines are caught up in all three clouds, such as the red case.
In order to investigate the correlation between a case-level characteristic and a case's participation in a given high-level path, we compare the participating cases only to those cases which visit the same locations in the process where the high-level behavior was observed.
In the example from Fig. \ref{fig: intro}, these would be segments $(a,b)$, $(b,c)$, and $(c,d)$ in this order.
Hence, in this work, we define the ``non-participating'' cases to be those case which are not participating, but could have participated from a control-flow perspective.
\section{Related Work}
Many of the recently developed process mining techniques acknowledge that the progress of cases in a process is influenced by the coexistence of other cases with which they must share process capacities. 
In \cite{zahra-causal}, the authors aim to improve the rate of positive outcomes in the process.
They propose appropriate interventions on changeable case aspects after having identified which treatments have a high causal effect on which case types.
This idea is taken further in \cite{zahra-prescriptive} where the appropriate time for applying a given time reducing intervention on a running case is determined.
This decision is based on the causal effect of the intervention which, in turn, includes
the number of active cases as an additional feature in the learning process.
Results in \cite{queue} show that the prediction of case delay at a certain activity is improved when the model is either a transition system whose state space is extended with system load information, or the model is based on queueing theory.
The method proposed in \cite{congestion-graphs} shows how information regarding workload and resource availability can be extracted from raw event data and encoded into congestion graphs. 
From these congestion graphs, congestion-related features can be extracted which are then used for predicting the time until next activity. 
Inter-case dependencies are also acknowledged in \cite{arik-chiara} where the current case prediction also factors in the predictions of the cases coexisting with the current one.
All these approaches acknowledge that process instances do not run in isolation and as such, integrating dedicated features which capture congestion to train time prediction models improves their accuracy.
While many of the high-level patterns we analyze in this work relate to process congestion, we do not encode our observations as additional features describing our running cases for prediction purposes.
Instead, each sequence of recurring outlier observations represents an explicit variant of high-level behavior which is a trait of the process itself.
We then look back into the ``low-level'' cases which were part of these observations to reason on whether a specific case type is over- or underrepresented in this group.

The performance spectrum \cite{fine-grained} clearly showed that processes---even within the same segment---exhibit non-stationary behavior which is not observable under aggregation.
The emerging patterns can reveal batching behavior \cite{eva-batch} which can have an influence on performance.
In \cite{eva-time}, the authors use visual analytics techniques based on the performance spectrum to demonstrate how errors in remaining time prediction are reduced when information on batching behavior is encoded in the learning process.
Our method conceptualizes many of the patterns that can be seen in the performance spectrum---including batching---through dedicated events, which can then be mined for further automated analysis.

Several recent methods have been developed which analyze how resources handle tasks from concurrently active cases within a process.
In \cite{priority}, the authors provide insights into how resources prioritize their work by employing specific queueing disciplines when processing individual tasks.
In \cite{batch-detection}, the authors detail the detection of batching behavior not only at the level of individual tasks but also across several linked activities.
Moreover, the approach outlined in \cite{batch-config} considers various batch behaviors concerned with multiple perspectives such as activities, resources, and data perspectives as well as allowing for batch detection even when batching is temporarily interrupted.
In \cite{profiles}, the authors introduce an enhanced resource profiling technique that considers not only the executed activities, but also the context (duration, case attributes) as well as multitasking.
In our work, we incorporate the resource dimension within the high-level events that describe outlier observations concerning batching and workload in specific process segments.
However, these observations are local and temporary and their emergence becomes relevant only in relation to the types of cases involved. 

In \cite{cross-case}, the authors introduce the concept of contextual association, wherein a group of cases exclusively exhibits concept drift whenever a shared object is subject to a change.
In our work, cases which give rise to high-level behavior are contextually associated due to their shared location and time in which the behavior is observed.
In this scenario, the term ``context" solely represents the \emph{coordinate} of a temporary observation in the process. 

The idea of conceptualizing outlier behavior related to load and delays as events themselves was first introduced in \cite{zahra}.
In that work, the emerging \emph{system-level events} from a Baggage Handling System (BHS) were connected based on time and place proximity.
The resulting cascades revealed how undesired system-level behavior arose and propagated throughout the BHS.
Similar work extending this idea was done in \cite{error-cascade} where DBSCAN is used to find frequent sequences of anomalies arising at the system-level.

The method we propose in this paper fits in the \emph{high-level event mining} framework we introduced in \cite{hlem}.
Each high-level event consists of the type of behavior detected, the entity involved and the time of detection.
In this work, we extend the types of high-level events that can be observed at the segment-level and propose a more refined way of correlating them.
Ultimately, we take a look at the underlying process instances and explore whether associations exist.
\section{Preliminaries}
\begin{definition}[Power set, Sequence, Suffix]\label{def:sequence}
    Given a set $A$, $\mathcal{P}(A)$ is the \emph{power set} of $A$ and $A^*$ are the finite sequences over $A$.
    For any $s,s' \in A^*$, we say $s'=\langle a_1',...,a_m' \rangle$ is a \emph{suffix} of $s=\langle a_1,...,a_n \rangle$ (denoted $s' \preceq s$) if and only if there is some $i \in \{0,...,n-m\}$ such that for $j=1,...,m$ it holds that $a'_{j} = a_{i+j}$.
\end{definition}
\begin{definition}[Events, Event log]\label{def:event log}
    $\mathcal{U}_{\mi{ev}}$ is the \emph{universe of events} and $\mi{Act}$, $\mi{Case}$, $\mi{Res}$ are the sets of \emph{activity names}, \emph{case identifiers} and \emph{resource names}, respectively.
    $T$ is the totally ordered set of \emph{timestamps}.
    $L=(E,\mi{Attr},\pi)$ is an \emph{event log} where $E\subseteq\mathcal{U}_{\mi{ev}}$ is a finite set of events, ${\{\mi{act}, \mi{case}, \mi{res}, \mi{time}\} \subseteq \mi{Attr}}$ is a set of attribute names and $\pi \in E \times \mi{Attr} \not \to \mi{Val}$ a (partial) function that assigns each event $e$ a value $\pi(e,\mi{att})$ or is undefined (written $\pi(e,\mi{att})\!\!=  \perp$).
    For any $e \in E$, $\pi(e,\mi{act}) \in \mi{Act}$, $ \pi(e,\mi{case}) \in \mi{Case}$, $\pi(e, \mi{res}) \in \mi{Res}$, and $\pi(e,\mi{time}) \in T$.
\end{definition}
For any attribute $\mi{att} \in \mi{Attr}$, we write $\mi{att}(e)$ instead of $\pi(e,\mi{att})$ when the event log is clear from the context.
Moreover, we assume that any two events of the same case never have identical timestamps.
\begin{definition}[Traces, Steps, Segments]\label{def:steps}
    The cases of an event log $L=(E, \mi{Attr}, \pi)$ are $\mi{cases}(L)=\{\mi{case}(e) \mid e \in E\}$.
    For any case $c \in \mi{cases}(L)$ with corresponding event set $E_c = \{e \in E \mid \mi{case}(e) = c\}$, the \emph{trace} of $c$ is the sequence $\sigma(c)=\langle e_1,...,e_{|E_c|} \rangle \in E_c^*$ containing all events from $E_c$ ordered by time,
    i.e., $\forall_{1 \leq i < j \leq |E_c|} \ \mi{time}(e_i) < \mi{time}(e_j)$.
    A \emph{step} is a pair of directly following events in a case in $L$.
    More precisely, the steps of $L$ are ${\mi{steps}(L)} = \{(e,e') \in E \times E \mid \exists_{c \in \mi{cases}(L)} \ \sigma(c) = \langle ...,e,e',... \rangle \}$.
    Moreover, we define $S(L)=\{(\mi{act}(e),\mi{act}(e')) \mid (e,e') \in \mi{steps}(L)\}$ as the \emph{segments} of $L$.
\end{definition}
A step is a pair of events which happened directly after each other in the same case.
A segment is a pair of activities that directly follow each other in the log.
\begin{definition}[Framing, Time Windows]\label{def:framing}
    A \emph{framing} is a function ${\phi \in T \rightarrow \mathbb{N}}$ mapping timestamps to numbers such that $\forall_{t_1, t_2 \in T} \ t_1 < t_2 \Rightarrow \phi(t_1) \leq \phi(t_2)$.
    Each $w \in \mi{rng}(\phi)$ represents time window  $\overrightarrow{w}=[w_{\mi{start}}, w_{\mi{end}}]$, where $w_{\mi{start}}= \mi{min} \{t \in T \mid \phi(t) = w\}$ and $w_{\mi{end}}= \mi{max} \{t \in T \mid \phi(t) = w \}$.
\end{definition}
Given an event log $L=(E,\mi{Attr},\pi)$ and a framing $\phi$, set $W_{L,\phi}=\{w \in \mathbb{N} \mid 
\mi{min}\{\phi(\mi{time}(e)) \mid e \in E\} \leq w \leq \mi{max}\{\phi(\mi{time}(e)) \mid {e \in E}\} \}$ contains all time windows of $L$ w.r.t. framing $\phi$.
Note that for any $e \in E$, $\phi(\mi{time}(e)) = w$ whenever $e$ occurred within $\overrightarrow{w}$.
\section{Method}
\subsection{Detecting High-Level Behavior Using High-Level Events}
The example in Sec. \ref{sec:example} illustrated three important components which comprise high-level behavior: the \emph{type} of behavior observed, the \emph{location} in the process where it emerges and the \emph{time} aspect related to it.
We call each pair of location and time information a \emph{coordinate}.
\begin{definition}[Coordinates]\label{def:co}
    Given log $L=(E, \mi{attr}, \pi)$ and window set $W_{L,\phi}$ w.r.t framing $\phi$, let $W_{L,\phi}^2=\{(w_1,w_2) \in W_{L,\phi} \mid w_1 \leq w_2 \}$.
    The set $\mi{CO}(L,\phi) = S(L) \times (W_{L,\phi} \cup W_{L,\phi}^2)$ contains the \emph{coordinates} of log $L$ w.r.t. $\phi$.
    Each coordinate $co=(s,\theta) \in CO(L,\phi)$ refers to a position in space (segment $s$) and time (window if $\theta  \in W_{L,\phi}$ and window pair if $\theta \in W_{L,\phi}^2$).
\end{definition}
Each outlier observation we considered in Sec. \ref{sec:example} emerged from a specific set of steps.
Which steps were involved in the observation depended on the type of behavior we were looking for.
%In that example, the \emph{type} of behavior observed was reflected in the cloud color.
For instance, the steps in the first cloud in Fig. \ref{fig: intro} represent the incoming load at segment $(a,b)$ within a particular time window.
Next, we conceptualize the colored clouds and the steps that are involved in them using \emph{high-level features}.
Each high-level feature consists of its type and a pattern.
One can think of the type being the color of the cloud and the pattern being the function which determines which subset of steps occurring in a given coordinate may give rise to that type of feature.
\begin{definition}[Pattern, Feature type]\label{def:pattern}
    Given a log $L=(E, \mi{attr}, \pi)$ and framing $\phi$, a \emph{pattern} is a (partial) function $\rho_{L,\phi} \in CO(L,\phi) \not \to \mathcal{P}(E \times E) \times \mathbb{R}$ which assigns a set of event pairs and a number to each coordinate of $L$ and $\phi$.
    %An \emph{evaluation} is a function $\mi{eval}_L \in \mathcal{P}(E \times E) \to \mathbf{R}$ which assigns a real value to a set of event pairs.
    %\mathcal{U}_{type}$ is the universe of types.
    A \emph{high-level feature} w.r.t. $L$ and $\phi$ is a pair $\mi{hlf} = (\mi{type}, \rho_{L,\phi})$ where $\mi{type} \in \mathcal{U}_{\mi{type}}$ is a feature type from the universe $\mathcal{U}_{\mi{type}}$ of feature types and $\rho_{L,\phi}$ is its pattern.
\end{definition}
In the remainder, given log $L$ and framing $\phi$, we write $\rho^{\mi{type}}_{L,\phi}$ to refer to the pattern of the high-level feature of type $\mi{type}$.

In this work, we consider feature types $\mi{enter}$, $\mi{exit}$, $\mi{workload}$, $\mi{handover}$, $\mi{batch}$, and $\mi{delay}$.
For each of these feature types, we now show how their patterns are determined.
Let $\mi{co}=(s,w) \in S(L) \times W_{L,\phi}$ be a coordinate from log $L$ with time windows from framing $\phi$.
Let $I_s = \{(e,e') \in \mi{steps}(L) \mid (\mi{act}(e), \mi{act}(e')) = s\}$ be the event pairs (steps) that traverse segment $s$ in the process and let $I_w=\{e \in E \mid \mi{time}(e) \in \overrightarrow{w}\}$ be the events that occur within time window $w$.
Feature type \emph{enter} is concerned with the steps that enter segment $s$ during $\overrightarrow{w}$ and thus $\rho_{L,\phi}^{\mi{enter}}(co) = (I^{\mi{enter}},\mi{val}) $ where $I^{\mi{enter}}=\{(e,e') \in I_s \mid e \in I_w\}$ and $\mi{val}=|I^{\mi{enter}}|$.
Feature type \emph{exit} is concerned with the steps that exit segment $s$ during $\overrightarrow{w}$ and thus $\rho_{L,\phi}^{\mi{exit}}(co) = (I^{\mi{exit}},\mi{val}) $ where $I^{\mi{exit}}=\{(e,e') \in I_s \mid e' \in I_w\}$ and $\mi{val}=|I^{\mi{exit}}|$.
Feature type \emph{workload} is concerned with the steps that exit segment $s$ during $\overrightarrow{w}$ for which it is the same resource executing both activities of $s$. 
Thus, $\rho_{L,\phi}^{\mi{workload}}(co) = (I^{\mi{wld}},\mi{val})$ where $I^{\mi{wld}}=\{(e,e') \in I_s \mid e' \in I_w \wedge \mi{res}(e)=\mi{res}(e')\}$ and $\mi{val}=|I^{\mi{wld}}|$.
Conversely, feature type \emph{handover} is concerned with the steps that exit segment $s$ during $\overrightarrow{w}$ for which there were two different resources executing the activities of $s$ (and so real work handover took place). 
Hence, $\rho_{L,\phi}^{\mi{handover}}(co) = (I^{\mi{hdo}},\mi{val})$ where $I^{\mi{hdo}}=\{(e,e') \in I_s \mid e' \in I_w \wedge \mi{res}(e)\neq\mi{res}(e')\}$ and $\mi{val}=|I^{\mi{hdo}}|$.

The time aspect of steps which are handled in batches refers to two time windows.
Let $(w,w') \in  W_{L,\phi}^2$ be a pair of time windows and let $\mi{co}=(s,(w,w')) \in S(L) \times W_{L,\phi}^2$ be a coordinate.
Let $I_{w,w'}=\{(e,e') \in \mi{steps}(L) \mid \mi{time}(e) \in \overrightarrow{w} \wedge \mi{time}(e') \in \overrightarrow{w'}\}$ be the set of event pairs (steps) where the first event occurs during $w$ and the second event occurs during $w'$.
Feature type \emph{batch} is concerned with the steps that enter segment $s$ during $\overrightarrow{w}$ and exit $s$ during $\overrightarrow{w'}$. 
Thus, $\rho_{L,\phi}^{\mi{batch}}(co) = (I_s \cap I_{w,w'}, |I_s \cap I_{w,w'}|)$.
Given a window distance $\delta \in \mathbb{N}$, the \emph{delay} w.r.t. to $\delta$ is concerned with the steps that together experience a similar delay which is at least $\delta$.
I.e., $\rho_{L,\phi}^{\mi{delay}}(co) = (I^{\mi{delay}},\mi{val})$ where $\mi{val}=|I^{\mi{delay}}|$ and $I^{\mi{delay}} = I_s \cap I_{w,w'}$ if $w' - w \geq \delta$ and $\emptyset$ otherwise.
\begin{definition}[High-level event]\label{def:hle}
    Let $L$ be an event log, $\phi$ a framing and $\mi{Type} \subseteq \mathcal{U}_{\mi{type}}$ a set of feature types.
    %Let $\mi{hlf}=(\rho, \mi{eval}) \in \mi{HLF}$ be a high-level feature and $r \in \mathbf{R}$ a threshold.
    Let $\mi{thr} \in \mi{Type} \times S(L) \to \mathbb{R}$ be a function assigning a threshold to any type-segment pair.
    We observe \emph{high-level event} ${h=(\mi{type}, \mi{co})} \in \mi{Type} \times \mi{CO}(L,\phi)$ with $\mi{co}=(s,\theta)$ if and only if $\rho_{L,\phi}^{\mi{type}}(\mi{co})=(I^{\mi{type}}, \mi{val})$ and $\mi{val} \geq \mi{thr}(\mi{type},s)$.
    Moreover, we call $C(h) \subseteq \mi{cases}(L)$ the \emph{cases of $h$} and for any $c \in \mi{cases}(L)$, it holds that $c \in C(h)$ if and only if there exists a step $(e,e') \in I^{\mi{type}}$ with $\mi{case}(e)=c$.
    The set $\mathcal{H}_{L,\phi,\mi{Type},\mi{thr}}$ contains all high-level events observed w.r.t. $L$, $\phi$, $\mi{Type}$, and thresholds from $\mi{thr}$.
\end{definition}
For any of the six feature types described above, a high-level event of type $\mi{type}$ is observed at coordinate $\mi{co}$ if and only if the number of event pairs that comprise the pattern related to $\mi{type}$ at that coordinate is higher than a given threshold.
Note that we propose the threshold to be determined based on both the feature type and the segment.
For instance, for any segment $s$, one observes a high-level event of type \emph{exit} at coordinate $(s,w)$ whenever the number of steps leaving $s$ during $w$ is above the $p$th percentile of all numbers of steps which leave $s$ in any given window throughout the process.

In the remainder of this work, we fix $L$, $\phi$, $\mi{Type}$ and $\mi{thr}$ and set \linebreak ${\mathcal{H} = \mathcal{H}_{L,\phi,\mi{Type},\mi{thr}}}$.
\subsection{Connecting High-Level Events}
\begin{figure}[t]
\minipage{0.32\textwidth}
  \includegraphics[width=\linewidth]{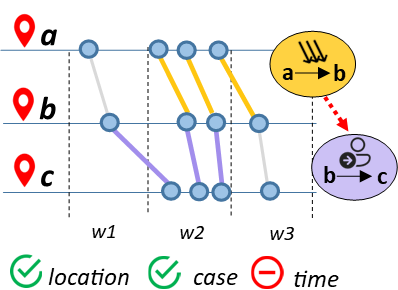}
\endminipage\hfill
\minipage{0.32\textwidth}
  \includegraphics[width=\linewidth]{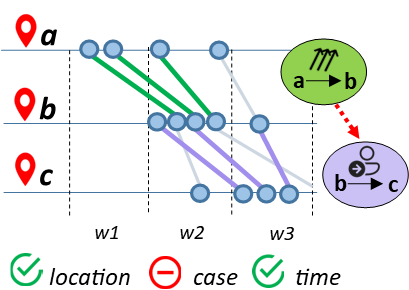}
\endminipage\hfill
\minipage{0.32\textwidth}%
  \includegraphics[width=\linewidth]{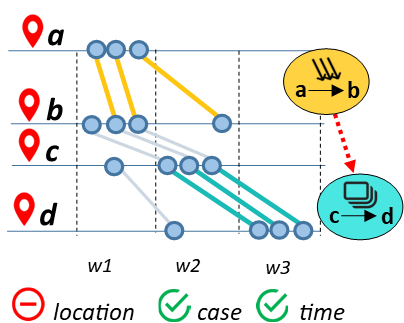}
\endminipage
 \caption{For $\lambda=0.5$, each figure shows an example where one overlap criterion from Def. \ref{def:overlap} is not satisfied. 
 In the left figure, high-level events $(\mi{enter},((a,b),w_2))$ and $(\mi{handover},((b,c),w_2)$ have no time overlap.
 In the middle figure, $(\mi{exit},((a,b),w_2))$ and $(\mi{handover},((b,c),w_3)$ have no sufficient case overlap.
 In the right figure, $(\mi{enter},((a,b),w_1))$ and $(\mi{batch},((c,d),(w_2,w_3)))$ have no location overlap.}
 \label{fig: overlap}
\end{figure}

The various high-level events observed throughout the process are not independent of each other.
As each high-level event relates to a time and place of occurrence, one can reason about time and space proximity.
Moreover, many of the steps in the patterns that give rise to them may belong to the same cases.
It seems natural to connect the three high-level events of types \emph{enter}, \emph{batch} and \emph{handover} observed in Fig. \ref{fig: intro} into a single \emph{episode} of high-level behavior.
That is because going from one high-level event to the next, one can notice that the cases involved have high overlap (1), and the first high-level event ``stops'' at the same place (2) and at the same time period (3) where the second one ``begins''.
We use the terms \emph{case overlap}, \emph{location overlap}, and \emph{time overlap} to refer to these connection criteria and we introduce them formally in this section.
\begin{definition}[Start Spread, End Spread]\label{def:spread}
    Let $h=(\mi{type}, \mi{co}) \in \mathcal{H}$ be a high-level event and let $\rho_{L,\phi}^{\mi{type}}(\mi{co}) = (I^{\mi{type}}, \mi{val})$.
    We refer to time period 
    \begin{align*}
        \mi{start}(h)= [ \mi{min} \{ \mi{time}(e) \mid (e,e') \in I^{\mi{type}}\}, \mi{max} \{ \mi{time}(e) \mid \{(e,e') \in I^{\mi{type}}\}]
    \end{align*}
    as the \emph{start spread} of $h$.
    Similarly, we refer to 
    \begin{align*}
        \mi{end}(h)= [ \mi{min} \{ \mi{time}(e') \mid (e,e') \in I^{\mi{type}}\}, \mi{max} \{ \mi{time}(e') \mid \{(e,e') \in I^{\mi{type}}\}]
    \end{align*}
    as the \emph{end spread} of $h$.
\end{definition}
In other words, given a high-level event related to segment $(a,b)$, the start spread covers the time period between the first and the last executions of $a$ from the steps in the corresponding pattern.
Similarly, the end spread covers the time period between the first and the last executions of $b$ from those same steps.
\begin{definition}[Overlap, Propagation]\label{def:overlap}
    Let ${h=(\mi{type},\mi{co})}, {h'=(\mi{type}',\mi{co}')} \in \mathcal{H}$ be two high-level events with $\mi{co}=(s,\theta)$ and $\mi{co}'=(s',\theta')$.
    Given some $\lambda \in [0,1]$, we say pair $(h, h')$ has \emph{case overlap} w.r.t. $\lambda$ 
    (denoted $h \overset{\mi{case}}{ \leadsto_{\lambda} } h'$) 
    if and only if 
    $\frac
    {|C(h_1) \cap C(h_2)|}
    {|C(h_1) \cup C(h_2)|}
    \geq \lambda. $
    We say pair $(h, h')$ has \emph{location overlap} 
    (denoted $h \overset{\mi{loc}}{ \leadsto } h'$) 
    if and only if $s=(a,b)$, $s'=(a',b')$ and $b=a'$.
    Moreover, we say pair $(h, h')$ has \emph{time overlap} 
    (denoted $h \overset{\mi{time}}{ \leadsto } h'$) 
    if and only if either $\mi{end}(h_1) \subseteq \mi{start}(h_2)$ or $\mi{start}(h_2) \subseteq \mi{end}(h_1)$.
    Ultimately, we say there is \emph{propagation} from $h$ to $h'$ w.r.t. $\lambda$ (denoted ${h \leadsto_{\lambda} h'}$) if and only if the pair $(h, h')$ has case overlap w.r.t. $\lambda$, location overlap and time overlap.
    More precisely:
    \begin{align*}
        \forall_{h,h' \in \mathcal{H}} \ {h \leadsto_{\lambda} h'} \ \Leftrightarrow \ 
    h \overset{\mi{case}}{ \leadsto_{\lambda} } h' \ \wedge \ 
    h \overset{\mi{loc}}{ \leadsto } h' \ \wedge \
    h \overset{\mi{time}}{ \leadsto } h'.
    \end{align*}
\end{definition}
%
% %
Two high-level events have time overlap whenever the end spread of the first one is contained in the start spread of the second one or the other way around.
Fig. \ref{fig: overlap} depicts examples of high-level event pairs that satisfy two of the overlap criteria, but not the third one.
%
% % 
% %
Whenever a pair of high-level events are close in time and space and their cases overlap sufficiently, we assume that their observations are correlated and we say that the first high-level event propagates to the second one.
Subsequent pairs of high-level events for which propagation occurs can lead to sequences which we call \emph{high-level episodes}.
\begin{definition}[High-level episode]\label{def:episode}
    Given high-level event set $\mathcal{H}$ and some case overlap threshold $\lambda$, any sequence of high-level events $\varepsilon = \la h_1,...,h_n \ra \in \mathcal{H}^*$ creates a \emph{high-level episode} if and only if
    $\forall_{1 \leq i < n} \ h_i \leadsto_{\lambda} h_{i+1}$ and $\bigcap_{h \in \varepsilon} C(h) \geq \lambda$.
    The set $\mathcal{E}(\mathcal{H}, \lambda)$ contains all such high-level episodes.
\end{definition}
In order to reason about recurring behavior, we abstract from the time component describing the high-level event and focus instead only on the type and location of the corresponding observation (the \emph{high-level activity}).
Moreover, we lift this concept to episodes and call the projection of an episode onto its high-level activities a \emph{high-level path}.
\begin{definition}[High-level path]\label{def:path}
   Let $\mathcal{H}$ be a set of high-level events and $\lambda \in [0,1]$.
   For any $h=(\mi{type}, \mi{co}) \in \mathcal{H}$ with $\mi{co}=(s,\theta)$, the \emph{high-level activity} of $h$ is $\mathbin{{h}{\uparrow}} = (\mi{type}, s)$.
   %Similarly, we lift episodes to \emph{high-level paths} by projecting the individual high-level events onto the high-level activities.
   For any episode $\varepsilon = \la h_1,...,h_n \ra \in \mathcal{E}(\mathcal{H}, \lambda)$, the sequence $\mathbin{{\varepsilon}{\uparrow}} = \la \mathbin{{h_1}{\uparrow}},...,\mathbin{{h_n}{\uparrow}} \ra$ is its corresponding \emph{high-level path}.
   Multiset $P(\mathcal{H}, \lambda) = [\mathbin{{\varepsilon}{\uparrow}} \mid \varepsilon \in \mathcal{E}(\mathcal{H}, \lambda)]$ contains all such high-level paths.
\end{definition}
Note that while each episode is unique because the high-level events are unique, the high-level paths may be recurring for different episodes.
It is for these recurring paths that we want to investigate the correlation with the properties of the cases involved.
\subsection{Case Participation in High-Level Behavior}
The participating cases of a given high-level path are those which are involved in all high-level events of an episode that executes the corresponding path.
The non-participating cases are those which are not participating, but which traverse the process segments underlying the path throughout their process run.
\begin{definition}[(Non-)participating cases]\label{def:participation}
    Given high-level event set $\mathcal{H}$ and case overlap threshold $\lambda$, let $p = \la h_1,...,h_n \ra \in P(\mathcal{H}, \lambda)$ be a high-level path and for each $i \in \{1,..,n-1\}$, let $s_i=(a_i,a_{i+1})$ be the segment where the high-level event $h_i$ was observed.
    The \emph{participating cases} of $p$ are $C_p = \{c \in \mi{cases}(L) \mid \exists_{\varepsilon \in \mathcal{E}(\mathcal{H}, \lambda)} \ 
    \mathbin{{\varepsilon}{\uparrow}}=p \ \wedge \ c \in \bigcap_{h \in \varepsilon} C(h)\}$.
    The \emph{non-participating cases} of $p$ are $\overline{C_p}=\{c \in \mi{cases}(L) \setminus C_p \mid \sigma(c) = \la e_1,...,e_k \ra \ \wedge \ \la a_1,a_2,...,a_n\ra \preceq \la \mi{act}(e_1), ..., \mi{act}(e_k)\ra\}$.
\end{definition}
To measure the correlation of a case-level attribute and a high-level path, the set of cases $C_p \cup \overline{C_p}$ is additionally partitioned according to the chosen case attribute value (see Table \ref{table:chi}).
The correlation is then computed using the $\chi^2$ test of independence on these two partitions.
The $\chi^2$ test measures the difference between the observed and expected frequencies for each combination of the values of two categorical variables.
The null hypothesis states that there is no relationship between case participation in a given high-level path and the chosen case attribute.
We consider the correlation as being statistically significant, and thus reject the null hypothesis, if the corresponding $p$-value of the result is smaller than $0.05$. 
\begin{table}[h!]\label{table:chi}
\caption{Given some high-level path $p$, the participating and non-participating case sets $C_p$ and $\overline{C_p}$ are further split based on the chosen categorical attribute values (here: category 1 and category 2). The correlation between the attribute and the high-level path is computed using the $\chi^2$ test of independence on the row partition (the chosen case-level attribute) and on the column partition ((non-)participation in the high-level path).}
\centering
\begin{tabular}{| c | c | c | c |}  
\hline
\textbf{Case-level attribute} & \textbf{Participating $C_p$} & \textbf{Non-participating $\overline{C_p}$}\\ \hline
category 1          & $n_1$                        & $n_2$                                      \\ \hline
category 2          & $n_3$                        & $n_4$                                      \\ \hline
                    & $n_1+n_3=|C_p|$              & $n_2+n_4=|\overline{C_p}|$                 \\ \hline
\end{tabular}
\end{table}
\section{Evaluation}
To evaluate our method, we used the BPI Challenge 2017 log\footnote{\url{https://data.4tu.nl/articles/dataset/BPI\_Challenge\_2017/12696884}}, which corresponds to a loan application process performed in a financial institution.
Each case in this event log is an application.
The applications can result in being successful or unsuccessful.
Moreover, the duration of applications varies from less than 10 days to over 30 days.
In this section, we analyze which sequences of high-level activities are strongly associated with the case attributes \emph{outcome} and \emph{throughput time}.
In the following, we briefly describe the event log, the setup of our experiments and some general statistics over the detected high-level events.
Afterwards, we comment on some of the high-level paths which showed a statistically significant correlation with the outcome and the throughput time of cases.
The method together with the evaluation script is available as a Python implementation\footnote{\url{https://github.com/biankabakullari/hlem-framework}}.
\subsection{The BPI Challenge 2017 Event Log}
The Application log of the BPIC 2017 contains a total of 31509 applications from January 2016 to February 2017.
The general control-flow of an application can be described as follows: 
first, a request for a loan is made.
Then, the submitted application is assessed.
If a credit offer can be made, the bank composes the offer and sends it to the customer.
Some time later, a bank employee calls the customer to remind them about the offer and answer any possible questions.
The customer sends all necessary documents to the bank which in turn has to validate them.
If the documents are incomplete, the customer is informed by a bank employee that that they have to resend the documents.
If the documents are complete, the bank either composes another offer, or it makes the ultimate decision to either accept or deny the application.
The activities of this process are divided into three categories: \emph{Application State Changes} (preceded by A\_), \emph{Offer State Changes} (preceded by O\_), and \emph{Workflow Events} (preceded by W\_).
The workflow events additionally contain \emph{lifecycle information} (e.g. schedule, start, suspend, resume, complete, ate\_abort/withdraw).
As the steps where cases move from one state of a workflow event to another make up for a significant amount of waiting time and rework in the process, we classify workflow related activities using both the event type and its lifecycle information.
This results in activities like {W\_Call after offers|SUSPEND} or \linebreak {W\_Call after offers|RESUME}.
\subsection{Experimental Setting and General Statistics}
Before applying our method on the event log, we projected the traces onto the 43 most frequent segments.
We split the time scope of the event data onto 398 windows, each corresponding to exactly one day.
We evaluated the patterns related to feature types \emph{enter}, \emph{exit}, \emph{workload}, \emph{handover}, \emph{batch} and \emph{delay} throughout the entire coordinate space.
Moreover, for \emph{workload} and \emph{handover}, we only considered human resources.
Here, it means that we removed User\_1 which was a system resource, and considered only real employees of the bank (User\_2 to User\_149).
Each observation counted as a high-level event whenever the measured number in the corresponding pattern was at least as high as the $90$th percentile of all the values obtained for that same type-segment pair.
For delays, we chose a $\delta$ based on the $70$th percentile of the days it takes to traverse a particular segment.
This resulted in a total of 5298 high-level events.
Table \ref{table:hle} shows the absolute and relative frequencies of high-level events for each feature type, together with the number of distinct segments where that type of high-level events was observed.
One can notice how the activities related to application files being incomplete (W\_Call incomplete files), the communication with the customers (W\_Call after offers) and the application validation (W\_Validate application) are most often subject to high-level behavior.
\begin{table}[t]\label{table:hle}
\caption{The total number of observed high-level events in the loan application log. For each feature type, one can see the absolute and relative number of high-level events of that type, the number of distinct segments where those high-level events were observed and the segment where they were observed most often.}
\centering
%\begin{tabular}{| m{5em} | m{2cm}| m{2cm} | m{2cm} | m{5cm} |}  
\begin{tabular}{| c | c | c | c |}  
\hline
\begin{tabular}{@{}c@{}}\textbf{ feature } \\ \textbf{ type }\end{tabular} & \textbf{\# hle (\%)} & \begin{tabular}{@{}c@{}}\textbf{\# distinct} \\ \textbf{segments}\end{tabular} & \textbf{most frequent segment} \\ \hline
workload & 1103 (20,82 \%)                 & 36                  &
\begin{tabular}{@{}c@{}}(W\_Call incomplete files|schedule, \\ W\_Call incomplete files|start)\end{tabular} \\ \hline
%(W\_Call incomplete files|schedule, \newline W\_Call incomplete files|start) 
handover & 214 (4,04\%)                    & 10                  & 
\begin{tabular}{@{}c@{}}(W\_Call incomplete files|suspend, \\ W\_Call incomplete files|resume)\end{tabular}\\ \hline
enter    & 1394 (26,31\%)                  & 43                  & ({A\_Create Application}, A\_Submitted)\\ \hline
exit     & 1377 (25,99\%)                  & 43                  & (A\_Create Application, A\_Submitted)\\ \hline
batch    & 1048 (19,78\%)                  & 11                  & 
\begin{tabular}{@{}c@{}}(W\_Call after offers|suspend, \\ W\_Call after offers|ate\_abort)\end{tabular}\\ \hline
delay    & 162  (3,06\%)                   & 5                   & 
\begin{tabular}{@{}c@{}}(W\_Validate application|suspend, \\ W\_Validate application|resume) \end{tabular}\\ \hline
\end{tabular}
\end{table}

We connected the high-level events into episodes using a case overlap threshold of $\lambda=0.5$.
This generated 102060 episodes, which corresponded to 68538 distinct high-level paths.
For these paths, we investigated the correlation with the \emph{outcome} and \emph{throughput time} of cases.
\subsection{Outcome: Success Rate}
The \emph{success rate} refers to the number of times an application results in a positive outcome (customer accepts an offer and the loan is granted) divided by the total number of applications.
In our event log, a \emph{successful} case translates into its trace containing activity A\_Pending.
In total, 17228 (54.85\%) cases are successful and the other 12183 (45.15\%) cases are unsuccessful.
The latter are the cases where the loan is either denied by the bank or cancelled by the customer.

Next, we show four frequent high-level paths which showed a statistically significant correlation with the case success rate.
For the two possible outcomes of \emph{success}, a significant correlation is observed whenever $\chi^2 \geq 3.841$.

\begin{figure}[t]
\minipage{0.4\textwidth}
  \includegraphics[width=\linewidth]{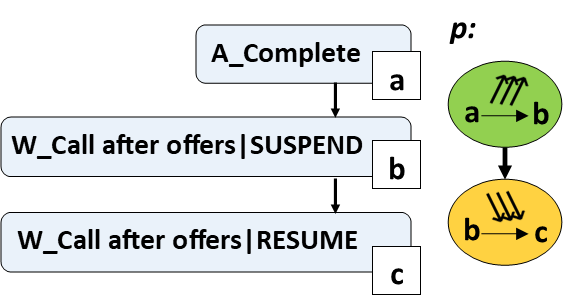}
  %\caption{A really Awesome Image}\label{fig:awesome_image1}
\endminipage \hspace*{0.5cm}
\minipage{0.45\textwidth}%
  \includegraphics[width=1\linewidth]{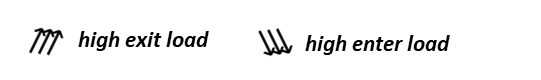}
  \begin{tabular}{| c | c | c |}  
\hline
\textbf{outcome} & \textbf{$C_p$} & \textbf{$\overline{C_p}$}\\ \hline
\textit{successful}       & 1087 (51,06\%)                & 11106 (53,48\%) \\ \hline
\textit{unsuccessful}     & 1042 (48,94\%)                & 9661 (46,52\%)   \\ \hline
Total: 22896             & $|C_p|=2129$               & $|\overline{C_p}|=20767$ \\ \hline
\end{tabular}
  %\caption{A really Awesome Image}\label{fig:awesome_image3}
\endminipage
\caption{The participating ($C_p$) and non-participating ($\overline{C_p}$) cases of the high-level path $p=\la (\mi{exit}, (a,b)), (\mi{enter},(b,c)) \ra$ where $a=$ A\_Complete, $b=$ W\_Call after offers|SUSPEND, and $c=$ W\_Call after offers|RESUME. 
This path was observed 14 times in the event log.
Here, $\chi^2\approx4,55$ and $p=0,0329$.}
\label{fig: path1}
\end{figure}

\begin{figure}[t]
\minipage{0.4\textwidth}
  \includegraphics[width=\linewidth]{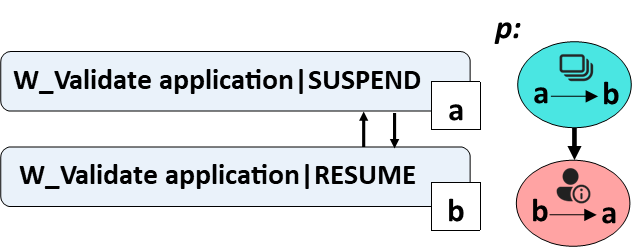}
  %\caption{A really Awesome Image}\label{fig:awesome_image1}
\endminipage \hspace*{0.5cm}
\minipage{0.45\textwidth}%
  \includegraphics[width=0.9\linewidth]{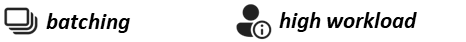}
 \begin{tabular}{| c | c | c |}  
\hline
\textbf{outcome}  & \textbf{$C_p$} & \textbf{ $\overline{C_p}$}\\ \hline
\textit{successful}       & 368 (72,73\%)                & 7160 (77,93\%) \\ \hline
\textit{unsuccessful}     & 138 (27,27\%)                & 2028 (22,07\%) \\ \hline
   Total: 9694                       & $|C_p|=506$                  & $|\overline{C_p}|=9188$ \\ \hline
\end{tabular}
  %\caption{A really Awesome Image}\label{fig:awesome_image3}
\endminipage
\caption{The participating ($C_p$) and non-participating ($\overline{C_p}$) cases of the high-level path $p=\la (\mi{batch}, (a,b)), (\mi{workload},(b,a)) \ra$ where $a=$ W\_Validate application|SUSPEND and $b=$ W\_Validate application|RESUME. 
This path was observed 10 times in the event log.
Here, $\chi^2\approx7,48$ and $p=0,0063$.}
\label{fig: path2}
\end{figure}

\begin{figure}[t]
\minipage{0.4\textwidth}
  \includegraphics[width=\linewidth]{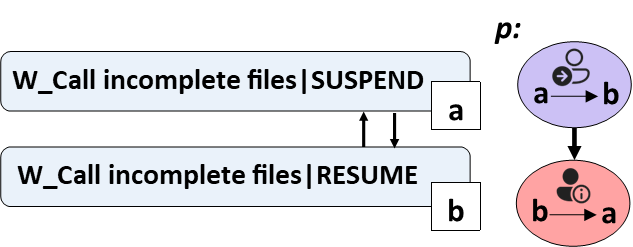}
  %\caption{A really Awesome Image}\label{fig:awesome_image1}
\endminipage \hspace*{0.5cm}
\minipage{0.45\textwidth}%
  \includegraphics[width=0.8\linewidth]{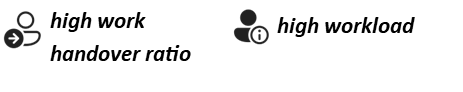}
 \begin{tabular}{| c | c | c |}  
\hline
\textbf{outcome} & \textbf{$C_p$} & \textbf{ $\overline{C_p}$} \\ \hline
\textit{successful}       & 852 (79,04\%)                & 10668 (85,05\%) \\ \hline
\textit{unsuccessful}     & 226 (20,96\%)                & 1875 (14,95\%)   \\ \hline
         Total: 13621                 & $|C_p|=1078$                 & $|\overline{C_p}|=12543$\\ \hline
\end{tabular}
  %\caption{A really Awesome Image}\label{fig:awesome_image3}
\endminipage
\caption{The participating ($C_p$) and non-participating ($\overline{C_p}$) cases of the high-level path $p=\la (\mi{handover}, (a,b)), (\mi{workload},(b,a)) \ra$ where $a=$ W\_Call incomplete files|SUSPEND and $b=$ W\_Call incomplete files|RESUME. 
This path was observed 16 times in the event log.
Here, $\chi^2\approx27,54$ and $p\approx1,53 \cdot 10^{-7}$.}
\label{fig: path3}
\end{figure}

\begin{figure}[t]
\minipage{0.4\textwidth}
  \includegraphics[width=\linewidth]{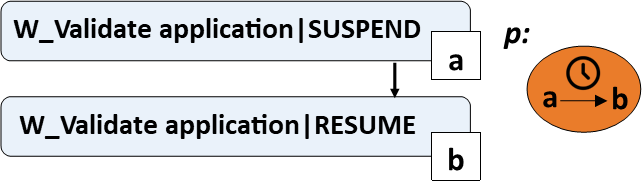}
  %\caption{A really Awesome Image}\label{fig:awesome_image1}
\endminipage \hspace*{0.5cm}
\minipage{0.45\textwidth}%
  \includegraphics[width=0.25\linewidth]{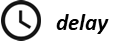}
\begin{tabular}{| c | c | c |}  
\hline
\textbf{outcome} & \textbf{$C_p$} & \textbf{ $\overline{C_p}$} \\ \hline
\textit{successful}       & 928 (72,22\%)                & 7822 (76,81\%) \\ \hline
\textit{unsuccessful}     & 357 (27,78\%)                & 2362 (23,19\%) \\ \hline
        Total: 11469                   & $|C_p|=1285$                 & $|\overline{C_p}|=10184$ \\ \hline
\end{tabular}
  %\caption{A really Awesome Image}\label{fig:awesome_image3}
\endminipage
\caption{The participating ($C_p$) and non-participating ($\overline{C_p}$) cases of the high-level path $p=\la (\mi{delay}, (a,b)) \ra$ where $a=$ W\_Validate application|SUSPEND and $b=$ W\_Validate application|RESUME. 
This path was observed 66 times in the event log.
Here, $\chi^2\approx13,28$ and $p=0,0002676$.}
\label{fig: path4}
\end{figure}

The path in Fig. \ref{fig: path1} shows that the success rate is lower for the cases which in large groups simultaneously go from having completed the application into the part where the bank initiates communication with them.
Moreover, it seems that for many cases in the process, the activities W\_Validate application and W\_Call incomplete files are first suspended, then resumed and afterwards suspended again.
Suspending after resuming seems to be associated with high resource workload (both paths in Fig. \ref{fig: path2} and \ref{fig: path3}). 
This high workload is preceded by batching behavior (Fig. \ref{fig: path2}) and high work handover ratio (Fig. \ref{fig: path3}).
Participation in both these high-level paths seems to also be negatively associated with the case success rate. 
Additionally, cases whose validation is resumed in batches with a long waiting time after the suspension (Fig. \ref{fig: path4}) seem to also show lower success rates than the cases whose validation is suspended and then later resumed with a shorter period in-between.
\subsection{Throughput Time}
The throughput time of a case is the time elapsed between the case's first and last event.
According to \cite{academic-bpic}, one can notice clear trends in the progress among the applications which take between 10-30 days to complete, and the applications which finish faster or pass the 30 days mark.
We use these throughput time categories to analyze the influence of our high-level paths on case duration.
In total, 7454 (23,73\%) cases finish in less than 10 days, 12963 (41,27\%) cases take between 10 and 30 days, and 10994 (35,00\%) cases spend longer than 30 days in the process.

Next, we show three high-level paths which showed a strong association with the case throughput time.
While it is unsurprising that high-level problems delay the progress of cases, our analysis reveals more detailed insights into what type of subsequent problems emerging at which process segments show particularly high association with the case duration.
For the three throughput time categories, a significant correlation is observed whenever $\chi^2 \geq 5.991$.

\begin{figure}[t]
\minipage{0.4\textwidth}
  \includegraphics[width=\linewidth]{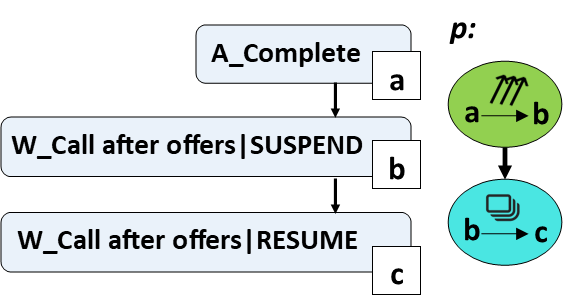}
  %\caption{A really Awesome Image}\label{fig:awesome_image1}
\endminipage \hspace*{0.5cm}
\minipage{0.45\textwidth}%
  \includegraphics[width=1\linewidth]{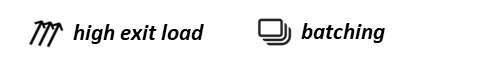}
\begin{tabular}{| c | c | c |}  
\hline
%\begin{tabular}{@{}c@{}}\textbf{throughput} \\ \textbf{time}\end{tabular} 
\textbf{thr. time}& \textbf{$C_p$} & \textbf{$\overline{C_p}$}\\ \hline
\textit{$\leq$ 10 days}       & 273 (13,75\%)                & 3860 (18,46\%)   \\ \hline
\textit{10-30 days}           & 848 (42,72\%)                & 8984 (42,96\%)   \\ \hline
\textit{$\geq$ 30 days}       & 864 (45,53\%)                & 8069 (38,58\%)  \\ \hline
                Total: 22898              & $|C_p|=1985$                 & $|\overline{C_p}|=20913$  \\ \hline
\end{tabular}
  %\caption{A really Awesome Image}\label{fig:awesome_image3}
\endminipage
\caption{The participating ($C_p$) and non-participating ($\overline{C_p}$) cases of the high-level path $p=\la (\mi{exit}, (a,b)), (\mi{batch},(b,c)) \ra$ where $a=$ A\_Complete, $b=$ W\_Call after offers|SUSPEND, and $c=$ W\_Call after offers|RESUME. 
This path was observed 15 times in the event log.
Here, $\chi^2\approx33,61$ and $p\approx5,04 \cdot 10^{-8}$.}
\label{fig: path5}
\end{figure}

\begin{figure}[t]
\minipage{0.4\textwidth}
  \includegraphics[width=\linewidth]{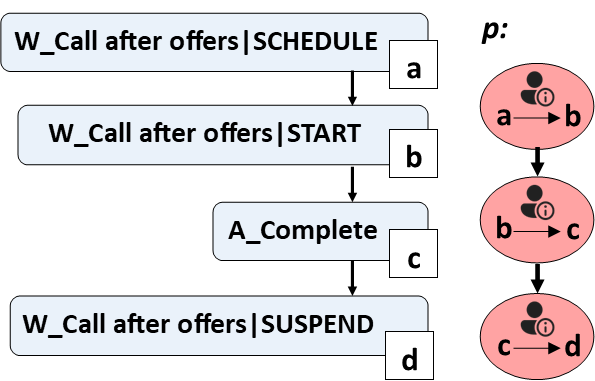}
  %\caption{A really Awesome Image}\label{fig:awesome_image1}
\endminipage \hspace*{0.5cm}
\minipage{0.45\textwidth}%
  \includegraphics[width=0.5\linewidth]{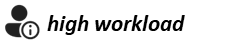}
\begin{tabular}{| c | c | c |}  
\hline
\textbf{thr. time} & \textbf{$C_p$} & \textbf{$\overline{C_p}$}\\ \hline
\textbf{$\leq$ 10 days}       & 224 (18,73\%)                & 6709 (23,20\%)   \\ \hline
\textbf{10-30 days}           & 550 (45,99\%)                & 12006 (41,53\%) \\ \hline
\textbf{$\geq$ 30 days}       & 422 (35,28\%)                & 10197 (35,27\%)  \\ \hline
     Total: 30108                         & $|C_p|=1196$                 & $|\overline{C_p}|=28912$\\ \hline
\end{tabular}
  %\caption{A really Awesome Image}\label{fig:awesome_image3}
\endminipage
\caption{The participating ($C_p$) and non-participating ($\overline{C_p}$) cases of the high-level path $p=\la (\mi{workload}, (a,b)), (\mi{workload},(b,c)), (\mi{workload},(c,d))\ra$ where $a=$ W\_Call after offers|SCHEDULE, $b=$ W\_Call after offers|START, $c=$ A\_Complete, and $d=$ W\_Call after offers|SUSPEND. 
This path was observed 15 times in the event log.
Here, $\chi^2\approx15,47$ and $p=0,000437$.}
\label{fig: path6}
\end{figure}

\begin{figure}[t]
\minipage{0.4\textwidth}
  \includegraphics[width=\linewidth]{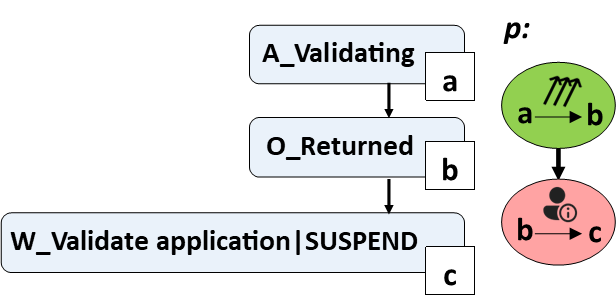}
  %\caption{A really Awesome Image}\label{fig:awesome_image1}
\endminipage \hspace*{0.5cm}
\minipage{0.45\textwidth}%
  \includegraphics[width=1\linewidth]{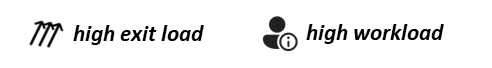}
\begin{tabular}{| c | c | c |}  
\hline
\textbf{thr. time} & \textbf{$C_p$} & \textbf{$\overline{C_p}$} \\ \hline
\textbf{$\leq$ 10 days}       & 448 (26,18\%)                & 5067 (30,28\%)  \\ \hline
\textbf{10-30 days}           & 1019 (59,56\%)               & 9542 (57,02\%)  \\ \hline
\textbf{$\geq$ 30 days}       & 244 (14,26\%)                & 2125 (12,70\%) \\ \hline
    Total: 18445                        & $|C_p|=1711$                 & $|\overline{C_p}|=16734$  \\ \hline
\end{tabular}
\endminipage
\caption{The participating ($C_p$) and non-participating ($\overline{C_p}$) cases of the high-level path $p=\la (\mi{exit}, (a,b)), (\mi{workload},(b,c))\ra$ where $a=$ A\_Validating, $b=$ O\_Returned, and $c=$ W\_Validate application|SUSPEND. 
This path was observed 14 times in the event log.
Here, $\chi^2\approx13,40$ and $p\approx0,000123$.}
\label{fig: path7}
\end{figure}

Similarly to Fig. \ref{fig: path1}, the path in Fig. \ref{fig: path5} shows that the cases which simultaneously transition from having completed the application into a part where the bank attempts to communicate with them in batches, also suffer longer throughput times.
One can notice that there are more cases that pass the 30 days' mark and less cases that finish in under 10 days from the group of participating cases than from the group of non-participating cases.
Moreover, having overloaded employees taking care of the communication process with the customers (the path in Fig. \ref{fig: path6}) also correlates with longer case throughput times (especially as the ratio of cases finishing in under 10 days decreases).
The path in Fig. \ref{fig: path7} covers the scenario when a case is validated, an offer is returned, but then the validating process has to be suspended.
It seems that for the cases which experience batching and high workload in this process part, the throughput time worsens.

To conclude, we noticed that participation in high-level behavior was negatively associated with both \emph{outcome} and \emph{throughput time} as the participating cases showed lower success rates and higher throughput times compared to the non-participating cases.
\section{Conclusion}
In this work, we aimed to explore the interplay between high-level problems in the process and the process instances which underlie them.
These problems were related to observations of high loads, busy resources, batching behavior and delays in particular locations in the process throughout different points in time.
We conceptualized each single outlier observation as a high-level event and we connected these high-level events into episodes whenever they were close enough in time, space, and when the cases giving rise to them were similar.
For a given sequence of (outlier) observations, we wanted to investigate whether the cases that participate in that high-level behavior differ significantly from the cases which do not.
For the comparison to be as meaningful as possible, the control group contained only the cases which were similar to the participating cases from a control-flow perspective.
Our experiments showed that for the loan application process, there were several examples of high-level behavior at particular segments which were negatively associated with the case outcome (loan application success rate) and throughput time.

While the same method can be applied w.r.t. any process property at the case-level, the discovered significance in the connection between the emerging process behavior and the process instances underlying it is bidirectional.
In future work, one could explore the cause-effect relationship behind these correlations.
For \emph{intrinsic} case properties (e.g., credit score of applicant), one could argue that it is the property itself which triggers certain kinds of high-level behavior.
For \emph{extrinsic} case properties (such as throughput time, or assigned resource for a specific task), a cause-effect discussion needs to consider the time when that property's value was set and when the high-level behavior emerged.
Moreover, we cannot exclude the presence of confounding variables, that is, process aspects which influence both case participation and the case characteristic considered.

A further improvement to the method could be in the automatic segment selection.
The experiments showed that some particular activities are run subsequently in an automatic way, so that reasoning about e.g., delays or work handover for these activity pairs makes less sense.
Moreover, one could extend the method with a way of evaluating and ordering the detected high-level behavior by how surprising or interesting it is for the process at hand.
Lastly, this method could be integrated in an interactive tool where the user selects the case property as well as high-level feature types and as a result, a list of most significant and interesting high-level behaviors w.r.t. that property is shown.

%
%\subsubsection{Acknowledgements} Please place your acknowledgments at
%the end of the paper, preceded by an unnumbered run-in heading (i.e.
%3rd-level heading).
\bibliographystyle{splncs04}
\bibliography{cascade_cases.bib}
%

%Author, F.: Article title. Journal \textbf{2}(5), 99--110 (2016)

%Author, F., Author, S.: Title of a proceedings paper. In: Editor,
%F., Editor, S. (eds.) CONFERENCE 2016, LNCS, vol. 9999, pp. 1--13.
%Springer, Heidelberg (2016). \doi{10.10007/1234567890}

%Author, F., Author, S., Author, T.: Book title. 2nd edn. Publisher,

%Author, A.-B.: Contribution title. In: 9th International Proceedings
%on Proceedings, pp. 1--2. Publisher, Location (2010)

%LNCS Homepage, \url{http://www.springer.com/lncs}. Last accessed 4
%\end{thebibliography}
\end{document}